# Biological Kerker effect boosts light collection efficiency in plants

Hani Barhom[1,2,=], Andrey A. Machnev[1,2,=], Roman E. Noskov[1,2,=,*], Alexander Goncharenko[3,4], Egor A. Gurvitz[5], Alexander S. Timin[6], Vitaliy A. Shkoldin[5,7], Sergei V. Koniakhin[7,8], Olga Yu. Koval[7], Mikhail V. Zyuzin[5], Alexander S. Shalin[5], Ivan I. Shishkin[1,2], and Pavel Ginzburg[1,2]

[1]Department of Electrical Engineering, Tel Aviv University, Ramat Aviv, Tel Aviv 69978, Israel
[2]Light-Matter Interaction Centre, Tel Aviv University, Tel Aviv, 69978, Israel
[3]Research Institute of Influenza, Ministry of Healthcare of the Russian Federation, Prof. Popova str. 15/17, St. Petersburg, 197376, Russia
[4]Peter The Great St. Petersburg Polytechnic University, Polytechnicheskaya str. 29, St. Petersburg, 195251, Russia
[5]Faculty of Physics and Engineering, ITMO University, Lomonosova 9, St. Petersburg 191002, Russia
[6] Research School of Chemical and Biomedical Engineering, National Research Tomsk Polytechnic University, Lenin Avenue 30, 634050 Tomsk, Russia
[7] St. Petersburg Academic University, 194021, St. Petersburg, Russia
[8] Institut Pascal, PHOTON-N2, Université Clermont Auvergne, CNRS, SIGMA Clermont, Institut Pascal, F-63000 Clermont-Ferrand, France

**Abstract:** Being the polymorphs of calcium carbonate ($CaCO_3$), vaterite and calcite have attracted a great deal of attention as promising biomaterials for drug delivery and tissue engineering applications. Furthermore, they are important biogenic minerals, enabling living organisms to reach specific functions. In nature, vaterite and calcite monocrystals typically form self-assembled polycrystal micro- and nanoparticles, also referred to as spherulites. Here, we demonstrate that alpine plants belonging to the *Saxifraga* genus can tailor light scattering channels and utilize multipole interference effect to improve light collection efficiency via producing $CaCO_3$ polycrystal nanoparticles on the margins of their leaves. To provide a clear physical background behind this concept, we study optical properties of artificially synthesized vaterite nanospherulites and reveal the phenomenon of directional light scattering. Dark-field spectroscopy measurements are supported by a comprehensive numerical analysis, accounting for the complex microstructure of particles. We demonstrate the appearance of generalized Kerker condition, where several higher order multipoles interfere constructively in the forward direction, governing the interaction phenomenon. As a result, highly directive forward light scattering from vaterite nanospherulites is observed in the entire visible range. Furthermore, *ex vivo* studies of microstructure and optical properties of leaves for the alpine plants *Saxifraga 'Southside Seedling'* and *Saxifraga Paniculata Ria* are performed and underlined the importance of Kerker effect for these living organisms. Our results pave the way for a bioinspired strategy of efficient light collection by self-assembled polycrystal $CaCO_3$ nanoparticles via tailoring light propagation directly to the photosynthetic tissue with minimal losses to undesired scattering channels.

*Corresponding author, e-mail: nanometa@gmail.com, phone: +972-3-640-6058
= These authors equally contributed

The polymorphs of calcium carbonate ($CaCO_3$), which include calcite, aragonite and vaterite, are known as the most abundant biogenic minerals. While there are many examples of naturally synthesized calcite and aragonite by living organisms or geological processes[1–3], vaterite appears in rarer but still extremely important cases. Specifically, vaterite has been detected in gallstones[4], human heart valves[5], green turtle eggshells[6], fish otoliths[7], mollusk pearls[8], and even in meteorites[9]. Apart from those relatively complex aggregations, vaterite can crystallize into polycrystalline micro- and nano-particles, also referred to as spherulites[10]. They are characterized by strong porosity, giving rise to high payload capacity. This extraordinary property along with natural biocompatibility makes vaterite spherulites a very promising platform for drug delivery[11–14].

Recently, vaterite- and calcite-producing cells have been discovered in several alpine plants from the *Saxifraga* genus[15]. They yield micron-scale vaterite or calcite particles on the leaf margins. This unusual biological strategy was shown to carry a functionality of paramount importance – light deflection into the adjacent leaf tissue, hypothetically improving efficiency of photosynthesis. Although qualitative explanation for this effect has been provided, the physical mechanism behind has remained obscure because scattering properties of such self-assembled biomineral nanoparticles have never been considered in detail.

Thus far, the studies of optical properties for calcite and vaterite nanoparticles have been predominantly focused on their birefringence[16]. In particular, rather large micron-size particles were considered owing to their relatively straightforward and repeatable fabrication protocols. Being highly anisotropic structures, vaterite and calcite particles have been used for demonstration of optomechanical rotations both in liquid solutions and vacuum[17–19]. However, systematic studies of sub-micron vaterite particles have not been performed owing to difficulties in developing reproducible self-assembly protocols and numerical complexity in description of their optical properties.

Here, we reveal the biological nature of the generalized Kerker effect, utilized by alpine plants belonging to the *Saxifraga* genus. Self-assembled calcium carbonate polycrystal micro- and nanoparticles, developed on the leaf margins, boost the light collection efficiency via mechanism of directional light scattering. We provide clear physical background behind the concept by comprehensive study of artificially synthesized subwavelength vaterite spherulites with dark-field spectroscopy and full-wave numerical simulations. In particular, spherulites with 440 nm, 490 nm and 540 nm sizes have been investigated. They demonstrate specific features in the dark-field scattering spectra, which are directly linked to the particle size and its internal structure. To elucidate these results, we employ finite element (FEM) numerical simulation taking into account internal nanostructure of spherulites and obtain quantitative agreement with the set of experiments. Field decomposition with respect to Cartesian multipoles reveals that vaterite spherulites meet the Kerker condition (i.e., in-phase scattering for electric and magnetic dipoles)[20] in a broad spectral domain. Similar behaviors have been recently found in man-made relatively high-index dielectric particles[21,22]. Moreover, we discover that vaterite spherulites also can meet the generalized Kerker condition[23], when almost equal in-phase scattering is obtained simultaneously from several electric and magnetic multipoles, leading to highly directive scattering. As a result, such particles, situated on the leaf surface, act as excellent matching elements significantly increasing efficiency of optical energy delivery to the photosynthetic tissue (mesophyll). This has been shown in numerical analysis and confirmed by *ex vivo* study of microstructure and optical properties for the leaves of *Saxifraga 'Southside Seedling'* and *Saxifraga Paniculata Ria*.

**Dark-field spectroscopy**

Dark-field spectroscopy is a powerful tool, which allows performing spectral analysis and imaging of particles with sizes below the diffraction limit[24,25]. Direct non-scattered light is removed by employing either Fourier filtering or inclined illumination of a sample. Here we implemented the latter approach, which is schematically depicted in Fig. 1 (a). This configuration is used to acquire and analyze the scattering from a particle, placed on a substrate. Positioning of the sample is achieved by using a motorized stage (MMP3, Mad City Labs). The studied objects can be visualized either in reflected light with fiber-coupled LED illumination or in the dark-field regime. The supercontinuum laser (YSL SC-Pro, 6W) was used as an excitation for the dark-field imaging and spectroscopy. The laser pulse picker was set to 1MHz repetition rate with 75% pump power output. The excitation was filtered using a 805 nm short-pass filter (Thorlabs) and coupled to a single-mode fiber. The side illumination was focused on the sample using 10X objective (Mitutoyo M Plan Apo NIR, NA = 0.28). The polarization state of the light incident on the sample was controlled by Glan-Thompson prism installed on the rotation mount. The scattered light was collected using 50X objective (Mitutoyo M Plan NIR, NA = 0.42). The angle of incidence on the sample surface was set to 65° from sample surface normal, thus the reflected excitation light is not directly coupled to collection optics. Under this angle of incidence, the illumination can be considered as a plane wave. An additional 1:1 telescope with knife-edge rectangular aperture (Ealing Optics) was introduced to the detection path in order to collect signal from a single particle and to increase signal-to-noise ratio. The collected light is split by 50:50 beam splitter and coupled to CMOS camera (Thorlabs DCC1240C) for imaging and fiber-coupled spectrometer (Andor Kymera 193i equipped with iDus 401 TEC-cooled CCD) for light analysis. The influence of the substrate was excluded by subtracting the background signal collected in a close vicinity of the analyzed particle from its scattering spectra. The reference signal was collected by replacing the sample with a spectralon target.

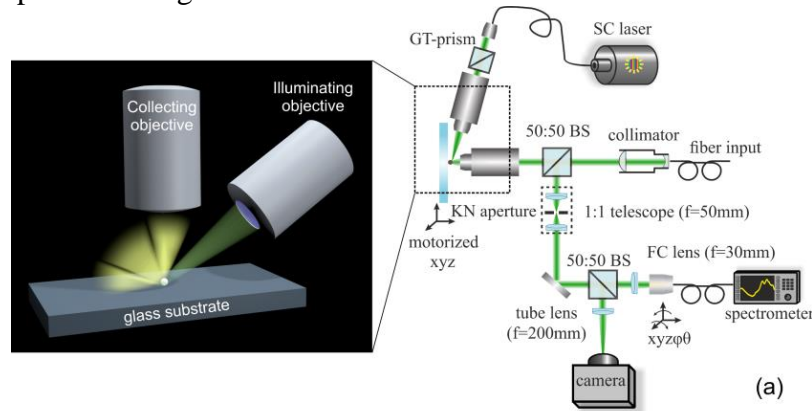

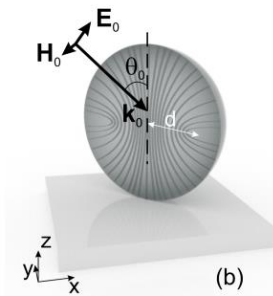
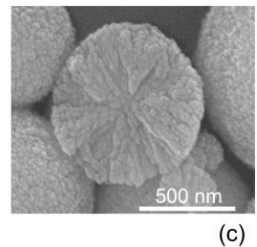

(a) (b) (c)

**Figure 1:** (a) Custom-made dark-field spectroscopy setup. TE- or TM-polarized supercontinuum laser beam is scattered by the nanoparticle on a substrate. Scattered light is detected by the collecting objective. Detailed layout of the experimental setup appears at the right of the panel (a). (b) Incidence of a TM-polarized light (the angle of incidence is $\vartheta_0 = 65^0$) on the vaterite spherulite. The inner structure of the particle is described as an array of densely packed co-focused hyperbolic fibers of single nanocrystal subunits, and $d$ is the distance between hyperbolas focus and the center of the particle. The dashed black line denotes the major optical axis of the spherulite. The particle is sliced in order to demonstrate its internal structure. (c) SEM image of the typical nanospherulite cross-section, showing size and distribution of vaterite monocrystals.

## Fabrication of vaterite spherulites

The vaterite particles have been prepared by using a modified protocol firstly described in Refs.[26,27] Calcium chloride dihydrate ($CaCl_2$, Sigma) and sodium carbonate ($Na_2CO_3$, Sigma) solutions were prepared in 83% ethylene glycol (EG, Sigma)/water mixture (5:1 in volume of EG:water). 2 mL of 0.33 M $CaCl_2$ solution was mixed with 386 µL of 0.33 M $Na_2CO_3$ under vigorous stirring at 1100 rpm for 25 min at room temperature. The reaction resulted in formation of sub-micron $CaCO_3$ particles. The formed particles were then washed twice with Milli-Q water by centrifuging (at 9000 rpm for 3 min) and dispersed in 1 mL of ethanol. Milli-Q water with a resistance greater than 18.2 MΩ cm$^{-1}$ was used for all experiments.

The typical scanning electron microscope (SEM) image, carried out on a Fei Quanta 200 Microscope with an acceleration voltage of 20 kV, appear in Fig. 1(c). Grain-based mesoporous structure, formed by self-assembled of nano-scale crystals, can be clearly identified. The average size of particles is around 500 nm in diameter with about 10% standard value deviation around this value as obtained from SEM images. One of the particles at the center of the image was spontaneously broken at the middle during the drying process and its internal structure appeared visible. So-called 'heap of wheat' structure, reported in literature[16,28], can be observed. Typical surface roughness of particles is about 30 nm and it is mostly dictated by the size of individual grains. Before measuring, the particles were dispersed and diluted in ethanol, and 50 µL drops were put onto the glass cover slip and dried at room temperature.

## Numerical model

Vaterite spherulites consist of single nanocrystal subunits that arrange bundles of fibers tied together at the center and spread out at the ends (so-called a "dumbbell" or "heap of wheat" model)[16,28]. Each subunit is a positive uniaxial monocrystal ($\varepsilon_{eo} > \varepsilon_o$) with $\varepsilon_o \approx 2.4$ and $\varepsilon_{eo} \approx 2.72$ (calcite has similar optical properties) [29,30]. The distribution of unity optical axes of such particles could be well approximated by the family of co-focused hyperboles, symmetrically rotated with respect to $z$-axis[16,31], as shown in Figs. 1 (b,c). The focal position $d$ can be controlled by the synthesis procedure. In the further analysis we set $d = r/2$ ($r$ is the particle radii), while this value can be adjusted by considering a specific fabrication protocol. Following the model presented in Refs.[31,32], we describe optical properties of such particles by the effective symmetric non-diagonal permittivity tensor whose all nine components are position-dependent. Those are obtained via rotations of the permittivity tensor of local monocrystals, oriented along the hyperbolic fibers, into a global system, associated with the macroscopic particle. For FEM simulations we use COMSOL Multiphysics and account for all details of the setup, including the microscopic permittivity tensor, angle of illumination, glass substrate, and numerical aperture (NA) of the collection optics.

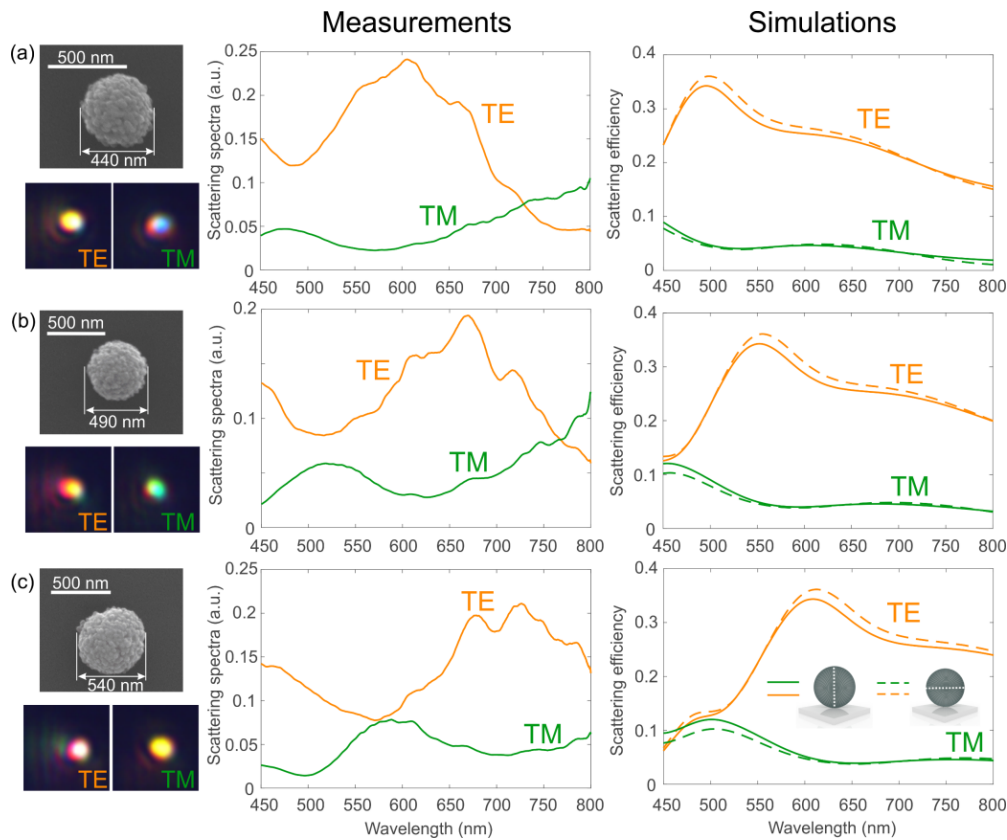

**Figure 2:** Experimental and theoretical dark-field scattering spectra for TE- and TM-polarized illumination. Left column - SEM and optical images correspond to the spectra on the right. Particles with 3 sizes are analyzed: (a) 440 nm, (b) 490 nm, and (c) 540 nm. Numerical simulations have been performed for 2 possible orientations of the major spherulite's optical axis with respect to the air-substrate interface. The corresponding spectra are denoted by solid and dashed curves (see inset in (c)).

**Results of dark-field measurements**

Comparison between measured and simulated dark-field scattering spectra for 440-nm, 490-nm, and 540-nm vaterite spherulites are shown in Fig. 2. Left column shows the SEM and optical images of the particles. After performing optical experiments, the particles were imaged with SEM. Identification of the samples was achieved by putting alignment marks on the glass substrate. First, optical images in different polarizations possess significant differences, visible by a naked eye – this is the manifestation for the strong sensitivity of the dark-field scattering spectra to the polarization of the illumination with respect to the substrate. Set of well-defined spectral features in the scattering can be also identified. Clear red-shift of the scattering peaks with the growth in the particle size can be observed both in the experiment and numerical analysis (panels from the top to the bottom). For TM polarization the incident angle is close to the Brewster condition. As a result, the substrate weakly reflects the light scattered by the particle to the imaging optics collecting cone. The forward scattering dominates the backward (as it will be shown in the next section), and one can observe quite a weak dark-field scattering signal in this case. Oppositely, for the TE polarization the substrate reflection is much stronger, which leads to well-pronounced peaks in the dark-field spectra associated with predominant scattering in the forward direction (see Fig. 4(c)).

In the experiment we did not control the orientation of the particles' major optical axes with respect to the substrate. To estimate impact of the internal spherulite anisotropy on the scattering, we performed numerical analysis for two uttermost orientations – major axis is either parallel or perpendicular to the substrate. The corresponding results appear in Fig. 2 and are indicated with dashed or solid lines. One can see that the orientation of the particle with respect to the substrate has only a minor effect on the scattering spectra. Thus, the spherulites scatter light almost equally for any polarization of the incident light, and the difference in spectral shapes for TE and TM cases is caused only by the Fresnel reflection of the forward-scattered light from the substrate.

The discrepancy between the experimental data and numerical fit mainly originates from two key factors. Spherulite surface roughness, which is of the same order as the size of single monocrystals (~30 nm), leads to diffuse light scattering, which was not taken into account in the numerical model. Additionally, slight ellipticity as well as small internal voids of spherulites, which are not visible on SEM images and can be only identified by breaking/slicing particles or with x-rays tomography[33], can play role.

**Irreducible Cartesian multipole decomposition**

There are a few possible multipole decompositions, suitable for analysis of electromagnetic scattering and radiation problems[34–36]. They have been applied to various systems and, obviously, provide identical scattering patterns. The choice of a particular decomposition is mainly related to results' interpretation and benefits for analysis of particular effects. Cartesian multipole decomposition is a more convenient and straightforward choice in the case of nontrivial scatterers. This representation, which is adopted here, allows separating contributions of high order irreducible tensors (basic multipoles) and explicit contributions of high-order toroidal moments up to toroidal electric octupole and toroidal magnetic quadrupole. This knowledge is essential for deriving generalized Kerker conditions in the case of broadband directive scattering from complex particles.

Multipole decomposition of the scattered field with 3 orders included is given by[37]:

$$E_{scat\_i} = \frac{k^2}{4\pi\varepsilon_0 R} e^{ikR} \left[ (n_i n_j - n^2 \delta_{ij}) \left( -\left( p_j + \frac{ik}{c} T_j^{(e)} + \frac{ik^3}{c} T_j^{(2e)} \right) + \frac{ik}{2}\left( \overline{\overline{Q}}_{jk}^{(e)} + \frac{ik}{c}\overline{\overline{T}}_{jk}^{(Qe)} \right) n_k \right.\right.$$
$$\left.\left. + \frac{k^2}{6}\left( \overline{\overline{O}}_{jkp}^{(e)} + \frac{ik}{c}\overline{\overline{T}}_{jkp}^{(Oe)} \right) n_k n_p \right) + \varepsilon_{ikj} n_k \left( -\frac{1}{c}\left( m_j + \frac{ik}{c} T_j^{(m)} \right) + \frac{ik}{2c}\left( \overline{\overline{Q}}_{jp}^{(m)} + \frac{ik}{c}\overline{\overline{T}}_{jp}^{(Qm)} \right) n_p + \frac{k^2}{6c}\overline{\overline{O}}_{jpl}^{(m)} n_p n_l \right) \right]. $$
(1)

We use exp(–*iωt*) for harmonic time-dependence and Einstein's summation notation, where vectors and tensors have lower-case indexes. The following parameters are defined: *k* is the vacuum wave-number, $R = |\mathbf{R}|$ the distance between the scatterer's centre and the obseravtion point, $\varepsilon_0$ the vacuum permittivity, $n_i = \dfrac{\mathbf{R}}{R}$; $p_j$ and $m_j$ basic electric dipole (ED) and magnetic dipole (MD), $\bar{\bar{Q}}^{(e)}{}_{jk}$ and $\bar{\bar{Q}}^{(m)}{}_{jp}$ electric and magnetic quadrupoles (EQ and MQ), $\bar{\bar{O}}^{(e)}{}_{jkp}$ and $\bar{\bar{O}}^{(m)}{}_{jpl}$ electric and magnetic octupoles (EO and MO). The toroidal moments are denoted with $T$. Superscripts in round brackets indicate their belonging to the corresponding basic multipole moments. Explicit expressions for multipole tensors can be found in[37]. Importantly, eq. (1) is derived for the scatterer in a homogeneous media. Strictly speaking, the presence of the substrate requires accounting for the secondary reflection of the light scattered by the particle from the substrate[38]. This contribution, however, is negligible as long as the substrate demonstrates a weak reflection towards the particle, which is a reasonable assumption here owing to low a contrast between glass and air refractive indexes.

Differential scattering cross-section is given by:

$$dP_{scat} = \frac{1}{2}\sqrt{\frac{\varepsilon_0}{\mu_0}}|\mathbf{E}_{scat}|^2 d\Omega. \qquad (2)$$

Normalization of $P_{scat}$ to the incident wave energy flux and the geometrical cross-section of the spherical scatterer gives the following expression for scattering efficiency:

$$\sigma_{scat} = \frac{2}{\pi r^2}\sqrt{\frac{\mu_0}{\varepsilon_0}}\frac{P_{scat}}{|\mathbf{E}_0|^2}, \qquad (3)$$

where $|\mathbf{E}_0|$ is the amplitude of the incident field.

We employ full-wave numerical simulations in COMSOL Multiphysics to solve the problem of light scattering from the vaterite spherulite on the glass substrate. Integration of the total scattered field in eq. (2) over the full solid angle gives the total scattering efficiency, while the dark-filed scattering efficiency is obtained by the integration over numerical aperture (NA) of the collecting objective. Results of the simulations, shown in Fig. 2, were obtained in this fashion.

Multipole decomposition is performed via direct integration of numerically calculated fields inside the particle. The accuracy of the decomposition is assessed by comparing the scattering efficiency obtained by straightforward integration of the scattered field and reconstructed from the multipole decomposition (eqs. (1) and (3)). Fig. 3(a) summarizes the convergence criteria – a slight discrepancy between blue and black lines underline weak secondary reflections from the glass substrate.

**Generalized Kerker condition for vaterite spherulites**

Total scattering cross-section and scattering pattern are governed by amplitudes of multipoles and relative phases between them. Different types of interference between multipoles can be achieved, depending on the configuration of the illumination field and electromagnetic properties of a scatterer. An example of such an interference phenomenon is a directional scattering, ensured by a condition, where electric and magnetic dipoles radiate in-phase (Kerker) or out-of-phase (anti-Kerker)[39–41]. Superior directivity can be achieved if high-order multipoles have proper phase and amplitude relations with respect to each other – this condition is referred to as a generalized Kerker effect[23,42]. Hereafter we will show, that this effect is achieved for the beforehand experimentally studied particles.

Multipole decomposition of the total scattering from the 440-nm vaterite particle excited with TE-polarized plane wave appears in Fig. 3 (a). While none of the multipoles demonstrates a strong resonant behavior, the overall scattering is built from many contributions. The case of the TM-polarized excitation is very similar and does not carry any additional critical information.

Figures 3 (b) and (c) show analysis of the far-field phases for the multipole contributions. We distinguish phases for the forward and backward directions with respect to the incident plane wave. This direction dependence can be clearly seen in eq. (1) through unit vectors $n_i$. For the longer wavelengths ED and MD dominate (Fig. 3(a)), and the relative phase shift between them is close to $\pi$ in the backward direction (Fig. 3(b)) and much less than $\pi$ in the forward direction. It means that the Kerker condition is nearly met over the entire observed spectrum. Moreover, higher order multipoles (MQ and EQ) start to contribute equally with ED and MD and demonstrate similar interference behavior at shorter wavelengths (approximately 500-550 nm). Such kind of high order multipole interference is referred to as one type from the family of the generalized Kerker effects[23]. Similar behavior has been recently revealed in completely different physical systems (see, e.g., [22,43] ).

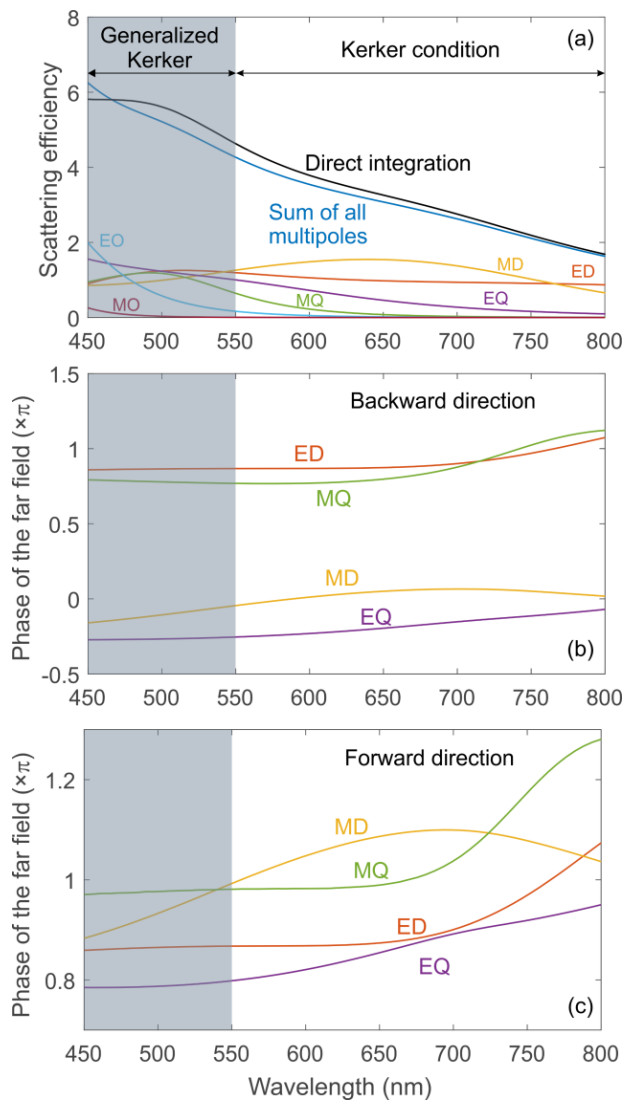

**Figure 3:** Analysis of scattering from 440-nm vaterite spherulite on the glass substrate excited by the TE-polarized plane wave. (a) Total scattering efficiency spectrum (scattering cross-section normalized by the particle geometric cross-section) and contributions to it from different Cartesian multipoles: ED – electric dipole; MD – magnetic dipole; EQ – electric quadrupole; MQ – magnetic quadrupole; EO – electric octupole; MO – magnetic octupole. Phases of the far-field terms corresponding to different multipoles in (b) backward and (c) forward directions with respect to the wavevector of an incoming wave.

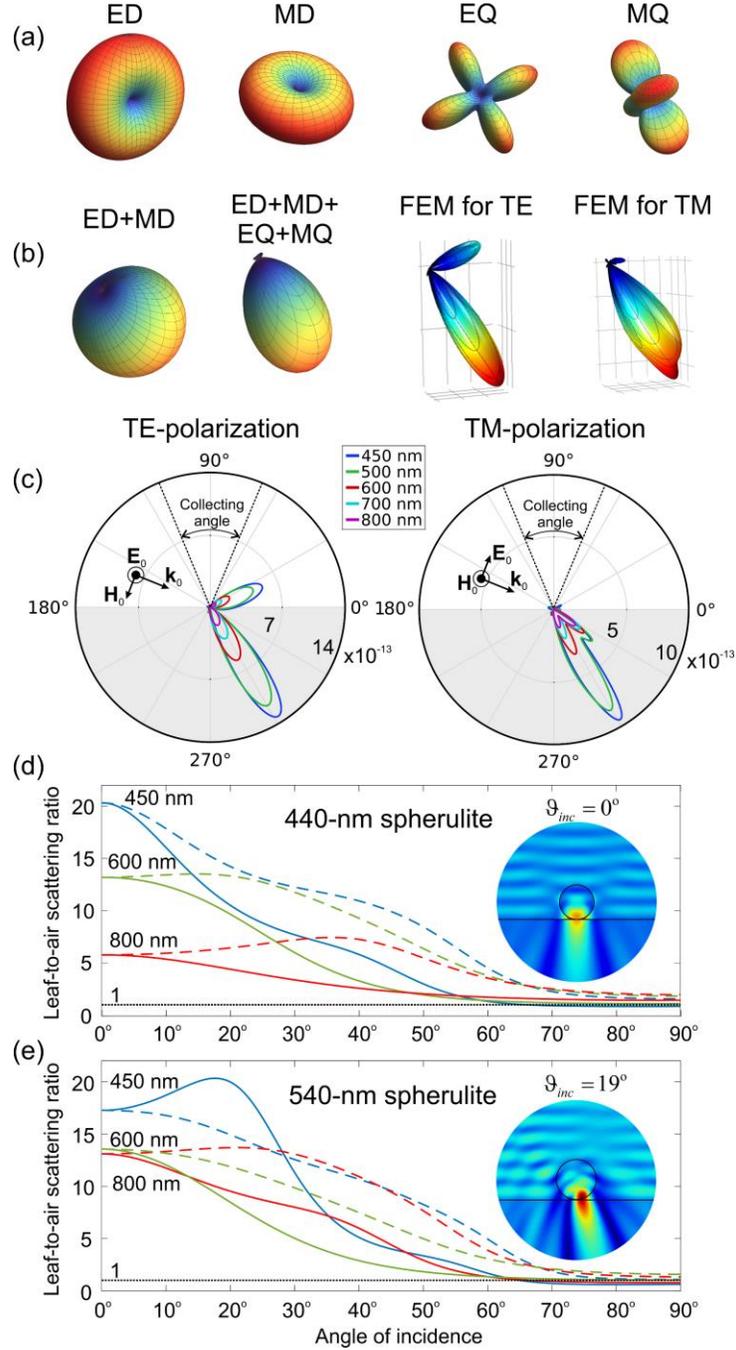

**Figure 4:** (a)-(c) Analysis of the Kerker effect for the 440-nm vaterite spherulite. (a) Far-field scattering patterns ($|E_{scat}|^2/|E_0|^2$) for isolated multipoles (ED, MD, EQ, MQ). (b) Their sums and direct numerical calculations (FEM) for TE- and TM-polarized incident light at λ = 500 nm. Multipole patterns do not account for the substrate contribution. (c) Scattering patterns ($|E_{scat}|^2/|E_0|^2$) at different excitation wavelengths (different colors) calculated numerically for TE- and TM-polarized incident light. The shadow areas denote the substrate. (d) and (e) Leaf-to-air scattering ratio vs the angle of plane wave incidence at λ = 450 nm, λ = 600 nm and λ = 800 nm for 440-nm and 540-nm vaterite spherulites. Full and dashed curves correspond to TE and TM polarizations, respectively. Insets show the distribution of the absolute value of electric field at λ = 450 nm and normal incidence in (d) and TE-polarized illumination at $\vartheta_{inc} = 19^o$ in (e). All simulations have been performed for the normal orientation of the major spherulite optical axis with respect to the air-substrate interface.

It is also instructive to analyze these Kerker and generalized Kerker effects at the level of scattering patterns. Figure 4 (a) demonstrates far-field scattering patterns for isolated multipoles. Their superposition with phase shifts extracted from full-wave simulations results in the forward scattering with quite a broad diagram at λ = 800 nm (ED+MD) and much more directive forward scattering at λ = 500 nm when 4 multipoles scatter nearly in-phase (ED+MD+EQ+MQ) (Fig. 4(b)). They are consistent with FEM simulations, and the only difference is that FEM accounts for the reflection from the substrate, while the multipole decomposition does not take it into account explicitly. Figure 4 (c) shows smooth growth in both total scattering power and scattering directivity as the wavelength decreases and the system is transforming from the Kerker to the generalized Kerker regime. Note that for λ = 450 nm electric octupole (EO) starts to provide a significant contribution to the scattering (Fig. 3(a)) and the directivity remains high, underlining that the generalized Kerker condition still works.

Remarkably, the total and the dark-field scattering cross-sections for 440-nm particle demonstrate slightly different spectral behavior (compare Fig. 2(a) and Fig. 3(a)). The total scattering cross-section monotonically increases as the wavelength gets shorter; while the dark-field signal features a pronounced peak at 500 nm. This stems from the fact that at shorter wavelengths the phase relationships between multipole components along with their individual contributions to the total scattering slightly vary. That gives rise to minor variations in the angular position of the maxima of the scattering diagrams at different wavelengths for forward scattered and reflected lobes. Specifically, for the TE-polarized excitation the reflected signal takes the maximum at the angle of $37^0$ for λ = 800 nm and $22^0$ for λ = 450 nm. Since the collecting solid angle is always fixed (Fig. 4 (c)), the shape of the dark-field spectral signal provides information about this drift in addition to the qualitative representation of the total scattering power.

Finally, to demonstrate the role of the Kerker effect in light collection of leaves, we calculate the ratio of light power scattered by 440-nm and 540-nm $CaCO_3$ spherulites into the leaf to the power scattered to the air (Figs. 4(d) and 4(e)). We took the refractive index of the leaf to be[44] 1.425 and take the Fresnel light reflection/refraction at the interface between the air and the leaf without a particle as a background electromagnetic field. This allows us to distinguish the contribution of the spherulite to the balance of the reflected and transmitted power from the conventional Fresnel reflection/refraction of an incident plane wave. The results are shown in Fig. 4(d). For the entire visible spectrum, both polarization and almost all angles of incidence, the $CaCO_3$ spherulites scatter much more optical power inside the leaf than back to the air, thus, acting as excellent matching elements. This is a straightforward consequence of the broadband Kerker effect. The effect is maximal for the normal or near-normal light incidence as a result of better transmission of such forward scattered waves through the air-leaf interface for both TE and TM polarizations (in accord to Fresnel equations). In contrast, at grazing light incidence the Fresnel reflectivity of the air-leaf interface rises, decreasing efficiency of light collection. Additionally, in the insets of Figs. 4 (d) and (e) one can observe generation of the nanojets, tightly focused light spots on the shadow side of the sphere, typical for homogeneous nanoparticles with a moderate refractive index[45].

It is important to note that since $CaCO_3$ spherulites are resonance-free, the Kerker effect for them is quite robust to variations in their geometric parameters. Light scattering by larger particles will involve more higher-order multipoles that may cause some changes in the leaf-to-air scattering ratio, but the general trend will not be altered, as shown in Figs. 4(d) and 4(e). Hence, our conclusions can be extended to $CaCO_3$ nanoparticles with a larger size and a variable shape.

### *Ex vivo* studies of *Saxifraga 'Southside Seedling'* and *Saxifraga Paniculata Ria*

In order to demonstrate functioning living organisms naturally producing self-assembled calcium carbonate particles, we perform *ex vivo* investigation of leaves for two alpine flowers: *Saxifraga 'Southside Seedling'* and *Saxifraga Paniculata Ria*. In these plants a special type of hydathodes (pores) secretes calcium carbonate on the leaf margins along with water through the pits (the process which is referred to as guttation) as shown in the SEM micrographs of a freshly cut leaf (Fig. 5). On the leaf margins of both samples the calcium carbonate forms the crystalline crust next to the hydathode pits that turns into aggregates of polycrystal particles whose size can vary in a wide range from several hundred of nanometers to several microns (Figs. 5(a) and 5(d)). We note that SEM images for *Saxifraga 'Southside Seedling'* demonstrate particles of two different kinds: ones made of comparatively large micron-scale prolongated monocrystal blocks (Fig. 5(b)) and other ones that contain very small nanocrystals ~ 30 nm (Fig. 5(c)). Such compositions are typical for calcite and vaterite particles, respectively. In contrast, particles generated by *Saxifraga Paniculata Ria* are more spherical (Figs. 5(e) and 5(f)).

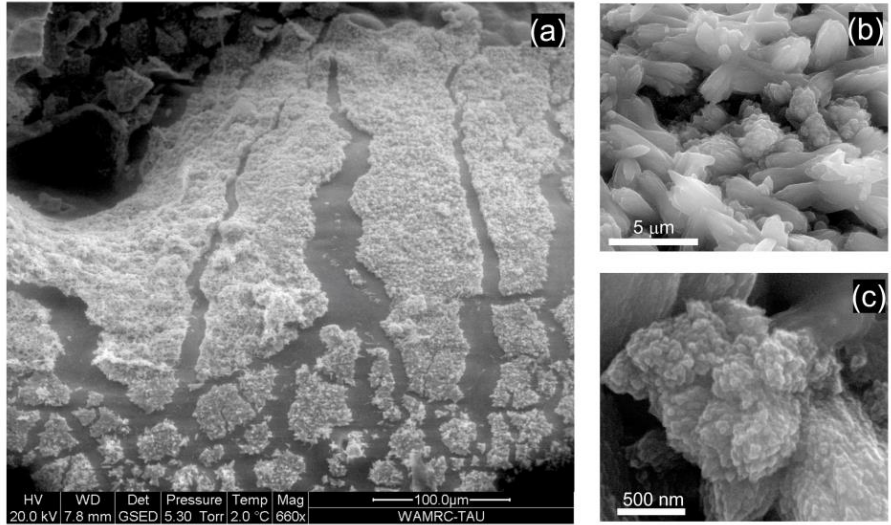

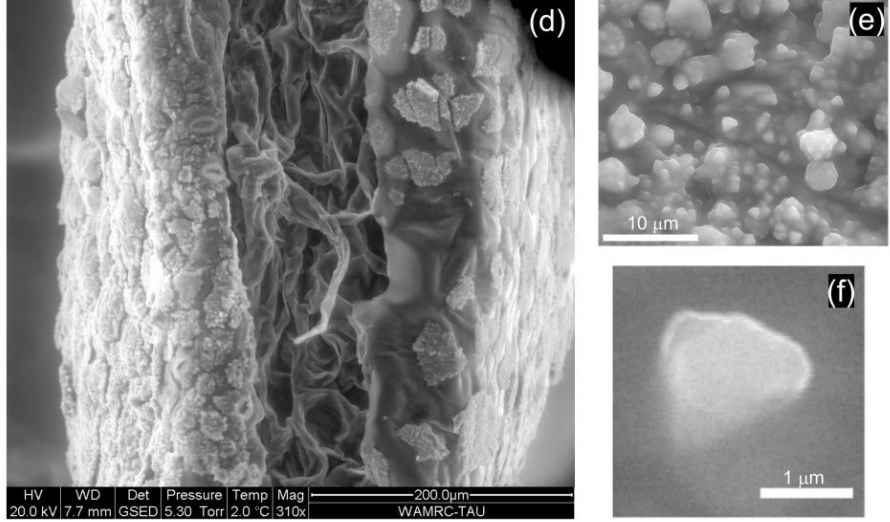

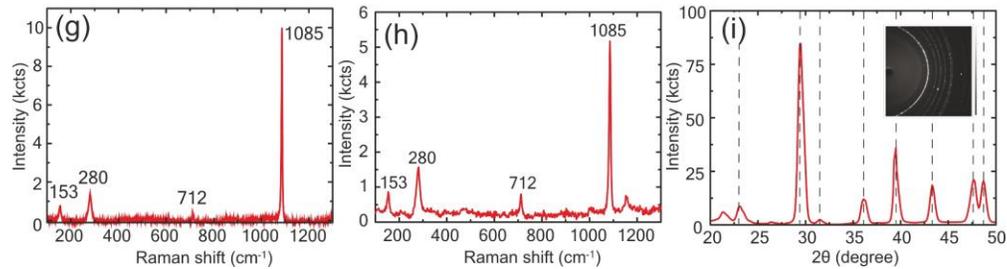

**Figure 5:** Scanning electron micrographs of freshly cut leaves for (a), (b), (c) *Saxifraga 'Southside Seedling'* and (d), (e), (f) *Saxifraga Paniculata Ria*. (a) Overview of a leaf margin with self-assembled aggregations of $CaCO_3$ micro- and nano-particles particles. Left upper corner: a hydathode pit that produces calcium carbonate along with water on the leaf surface as a result of guttation. (b) Zoomed area with typical particles. (c) Polycrystal nanospherulite. (d) Transversal structure of a leaf. The mesophyll (the primary site of photosynthesis) is situated in the middle between the leaf epidermis surfaces which hold aggregations of calcite micro- and nanoparticles zoomed in (e) and (f). The images were taken by FEI Quanta 200 microscope in the environmental mode (ESEM) with acceleration voltages of 20 kV, at the pressure of 5.3 Torr and the temperature +2.0 °C. (g) and (h)

show Raman spectra of the crystalline deposits for *Saxifraga Paniculata Ria* and *Saxifraga 'Southside Seedling'*, respectively. (i) XRD data of the crystalline deposits for *Saxifraga Paniculata Ria*. Dashed lines correspond to the calcite peaks. Inset in (i) demonstrates the crystalline deposit diffractogram.

To confirm our assumptions about the particles composition, we perform Raman spectroscopy by using the Olympus IX71 inverted microscope with a 50x objective coupled to the confocal spectrometer (Horiba LabRam HR). The sample was excited by the 532-nm laser with average power of 5 mW. No cover glass was present between the leaf and objective during measurement in order to avoid degradation of the signal. Raman spectrum demonstrates clear calcite peaks for both samples – at 153 $cm^{-1}$, 280 $cm^{-1}$, 712 $cm^{-1}$, and 1085 $cm^{-1}$ (Figs. 5(g) and 5(h)). The absence of Raman peaks, corresponding to vaterite, can be explained by very weak reflection of a signal from the areas where these particles are situated.

Additionally, the X-ray diffraction (XRD) analysis was carried out for *Saxifraga Paniculata Ria* by a Bruker Smart Apex Duo installation with a CuKα source and Apex 2D detector (Fig. 5(i)). Diffraction patterns were measured at the angle of $30^0$ between the normal to the detector surface and the incident X-ray direction. The 2D data was subsequently recalculated to the standard 2θ configuration. XRD also verifies the presence of calcite particles, and a set of rings in the diffractogram clearly shows polycrystalline nature of the deposits.

Interestingly, the SEM images of the dissected leaf margins (see Fig. 5(d)) allowed us to estimate if $CaCO_3$ particles are also present inside leaves. Such structures (also referred to as cystoliths) are known to be formed in certain families of plants and typically have the size ~10 - 100 μm[46,47]. Careful analysis of SEM images has shown that there is no $CaCO_3$ particles present inside leaves for the considered samples. Thus, the studied biomineral particles have different origin rather than cystoliths.

Our analysis has demonstrated generally similar organization and composition of $CaCO_3$ particles in both samples. Therefore, we, next, consider light-to-leaf interactions for *Saxifraga Paniculata Ria* only without loss of generality. Figure 6(a) shows the photograph of the sample leaf where the white crusty deposits represent biomineralized $CaCO_3$ aggregates. To observe the light refraction in these deposits, the crusty edge of the leaf was illuminated with the third harmonic (355 nm) of Yb solid-state femtosecond laser (TeMa, Avesta Project) with average power of 10 mW. The laser beam was focused on the leaf sample using a NUV-transparent objective (Mitutoyo M Plan Apo NUV 20x, 0.42NA) into a diffraction-limited spot and the resulting images were collected with a CCD camera. The unfiltered image of the leaf edge under excitation is shown in Fig. 6(b). The autofluorescence coming from the leaf tissue can be observed. In order to demonstrate its source of origin, we filtered out the image with 550 nm and 650 nm long-pass filters (Figs. 6(c) and 6(d)). This allowed us to collect a signal from different photosynthetic pigments in mesophyll (spongy tissue shown in Fig. 5(d)) whose fluorescence windows cover the range between 560-735 nm[48]. Hence, the capability of light deflection by biomineral deposits is confirmed by presence of chlorophyll autofluorescence far from the excitation spot.

It is instructive to study an internal part of the leaf participating in calcium carbonate particles generation. To this end, we employ confocal laser scanning microscopy of the dissected leaf (Figs. 6(e) and 6(f)). The images were collected by the Zeiss LSM 710 microscope with the 488-nm excitation laser. The red channel was acquired with the photomultiplier tube detection window set to 560-650 nm that corresponds to the chlorophyll fluorescence signal, whereas the green channel window was set to 500-550 nm that corresponds to the xylem tissue fluorescence signal. Chlorophyll fluorescence has grainy structure due to spongy constitution of mesophyll (Fig. 5(d)). The xylem forms the hydathodes transporting water and mineral salts from roots to stems and leaves so that one can observe a branchy array of such channels secreting calcium carbonate particles on the leaf surface (Fig. 6(e)). Moreover, the xylem appears to be included into calcium carbonate deposits (Fig. 6(f)). Although the role of this effect is unclear, one may hypothesize that xylem may participate in shaping of calcium carbonate particles and aggregates.

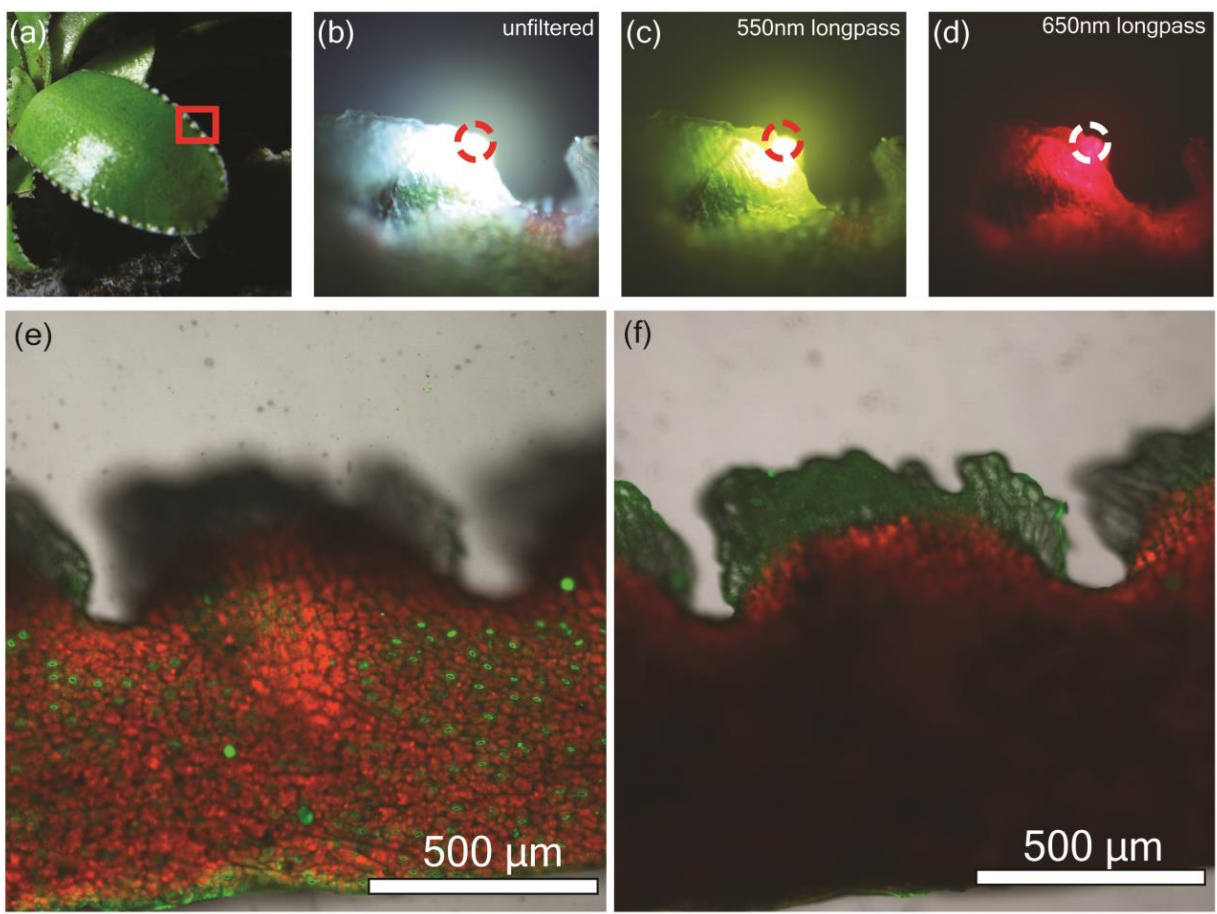

**Figure 6:** (a) Photograph of *Saxifraga Paniculata Ria*. Red square indicates aggregations of naturally produced $CaCO_3$ particles. (b) Unfiltered image of the leaf margin in the visible domain under the 355-nm illumination at the $CaCO_3$ aggregates on the leaf edge. The laser excitation spot is denoted by the dashed circle in (b), (c), and (d). (c) and (d) show images filtered with 550-nm and 650-nm long-pass filters, respectively. The yellowish and red colors correspond to the fluorescent signals from the photosynthetic tissue, mesophyll. (e,f) Confocal laser scanning microscopy images of the dissected leaf margin. The green and the red fluorescent signals originate from the xylem tissue and the mesophyll, respectively.

Finally, we note that in nature plants from the *Saxifraga* genus (including those studied in this paper) usually occur in rock cracks. Accounting for the fact that the $CaCO_3$ particles are synthesized only on the leaf margins (Fig. 6(a)), one may conclude that boosting light collection efficiency via the Kerker effect is necessary to look for sun rays (that can be in high demand in the rock cracks) and timely reorient leaves towards them, as was shown in Ref.[15].

**Conclusions and outlook**

We have investigated optical properties of self-assembled biomineral vaterite spherulites by means of the dark-field spectroscopy. Complex internal microstructure of the spherulites was introduced into numerical analysis through the non-diagonal and position-dependent tensor of the effective permittivity. Scattering characteristics of several submicron vaterite particles were considered with the help of irreducible Cartesian multipole decomposition, and significant contributions of high order multipoles were identified. In particular, we revealed multipole interference associated with broadband Kerker and generalized Kerker conditions, giving rise to directive forward light scattering. This type of phenomenon was demonstrated for the first time in a biogenic system. Our results manifest that calcium carbonate nanospherulites offer a promising playground for engineering light propagation due to broadband generalized Kerker effect that can be used to design bioinspired nanophotonic devices. Vivid biological examples of such concept are the alpine plants *Saxifraga 'Southside Seedling'* and *Saxifraga Paniculata Ria* which have evolved special cells synthesizing $CaCO_3$ micro- and nanoparticles to boost efficiency of light collection.

**Acknowledgement**

The research was supported in part by ERC StG 'In Motion' (802279), PAZY Foundation (Grant No. 01021248). Part of this work related to the synthesis of vaterite particles was supported by Russian Science Foundation, grant No. 19-73-30023. M.V.Z. thanks the President's Scholarship SP-1576.2018.4. S.V.K. acknowledges the support of the ANR "Quantum Fluids of Light" project (ANR-16-CE30-0021) and Project 16.9790.2019 of the Ministry of Science and Higher Education of Russian Federation. The authors acknowledge Dr. Zahava Barkay for help with ESEM of leaves, Dr. Gal Radovsky for help with Raman measurements, and Dr. Ivan Mukhin for helpful discussions.

**Authors contribution**

R.E.N. conceived the idea and initiated the project. H.B. and A.G. synthesized vaterite nanospherulites. A.M., I.I.S., H.B., and A.G. performed the dark field spectroscopy of samples. R.E.N. and A.E.G. did numerical analysis. R.E.N., A.M., and V.A.S. performed SEM of leaves. R.E.N., A.M, I.I.S., S.V.K., and O.Yu.K. performed Raman spectroscopy. S.V.K. and I.I.S. performed XRD analysis. A.S.T., I.I.S., and M.V.Z. performed confocal microscopy of leaves. A.S.S. contributed to Cartesian multipole decomposition. R.E.N. and P.G. wrote the manuscript and supervised the project.